\documentclass[%
preprint,
 amsmath,amssymb,
 aps,
]{revtex4-1}

\usepackage{graphicx}% Include figure files
\usepackage{dcolumn}% Align table columns on decimal point
\usepackage{bm}% bold math
\usepackage{braket}
\usepackage{amsmath}	% required for `\align' (yatex added)
\begin{document}

\preprint{APS/123-QED}

\title{Structure and decay of the pygmy dipole resonance in $^{\bm{26}}$Ne}% 

\author{M. Kimura}
\email{masaaki@nucl.sci.hokudai.ac.jp}
\affiliation{Department of Physics, Hokkaido University, Sapporo 060-0810, Japan}
\affiliation{Nuclear Reaction Data Centre, Faculty of Science, Hokkaido University,
Sapporo 060-0810, Japan}

\date{\today}% It is always \today, today,

\begin{abstract}
 The low-lying spectra of $^{24,25,26}{\rm Ne}$ and the structure of the pygmy dipole resonance
 (PDR) in  $^{26}{\rm Ne}$ have been theoretically studied by the antisymmetrized
 molecular dynamics (AMD) and its extended version called shifted-basis AMD. The calculated energy
 and strength of  the PDR reasonably agree with the observation, and the analysis of the wave
 function shows that  the PDR is dominated by neutron excitation coupled to the quadrupole excited
 core nucleus  $^{25}{\rm Ne}$, which explains the observed unexpected decay of PDR to the excited
 states of  $^{25}{\rm Ne}$. The large isoscalar component of PDR is also shown and the
 enhancement of the  core excitation in neutron-rich Ne isotopes is conjectured.

\end{abstract}

\pacs{Valid PACS appear here}% PACS, the Physics and Astronomy
\maketitle

%\tableofcontents

\section{introduction}
The low-energy electric dipole ($E1$) excitation which emerges well below the giant dipole
resonance (GDR) is called pygmy dipole resonance (PDR), and has attracted much interest in this
decade \cite{Paar2007a,Savran2013}. It has been expected that PDR can be a signature of a novel
type of excitation mode peculiar to unstable nuclei, in which the tightly bound inert core
oscillates against the  surrounding neutron skin
\cite{Mohan1971,Suzuki1990,VanIsacker1992}. Hence, the relationship between the strength of PDR
and the growth of neutron-skin in many isotope chains has been discussed by many authors
\cite{Inakura2009,Carbone2010,Reinhard2010,Piekarewicz2011,Piekarewicz2012}. In addition to this,
the PDR is expected to have a strong impact on astrophysical phenomena such as the rapid neutron
capture process, and constrains the equation of state of the neutron star matter 
\cite{Reinhard2010,Piekarewicz2011,Piekarewicz2012,Goriely1998,Goriely2004,Litvinova2009}. 

Among many observed PDR, that of $^{26}{\rm Ne}$ is the most intensively and detailedly
studied one. The experiment performed at RIKEN reported the PDR of $^{26}{\rm Ne}$ around 
$E_x=9$ MeV with the integrated $E1$ strength of $B(E1)=0.49\pm 0.16\ \rm e^2fm^2$ which exhausts
approximately 5\% of Thomas-Reiche-Kuhn (TRK) sum rule \cite{Gibelin2008}. Many theoretical
studies based on the quasi particle random phase approximation (QRPA) have been performed and
successfully described these observed properties, although the results range $E_x=6\sim 10$ MeV
and $5\sim10$ \% of TRK sum rule depending on the effective interactions used in the calculations 
\cite{Cao2005,Peru2007,Yoshida2008,Yoshida2008a,Inakura2011,Martini2011,Hashimoto2012,Inakura2013,Ebata2014,Peru2014}. 
At the same time, several QRPA calculations pointed out that the PDR of  
$^{26}{\rm Ne}$ is less collective and dominated by a limited number of neutron $1p1h$
excitations. For example, in Refs. \cite{Yoshida2008a,Martini2011}, it was shown that the PDR is
dominated by the $\nu(1s_{1/2}^{-1}1p_{3/2})$ and  $\nu(1s_{1/2}^{-1}1p_{1/2})$
configurations. However, at a glance, these $1p1h$ configurations look contradict to the observed
decay pattern of PDR. The dominance of the $1p1h$ configuration such as 
$\nu(1s_{1/2}^{-1}1p_{3/2})$ implies that the PDR primary decays to the ground state of 
$^{25}{\rm Ne}$ which has the  $\nu(1s_{1/2}^{-1})$ configuration relative to the ground state of
$^{26}{\rm Ne}$. On the other hand, experimentally, it was found that the PDR of $^{26}{\rm Ne}$
predominantly decays into the excited states of $^{25}{\rm Ne}$, not to the ground state
\cite{Gibelin2008}. This puzzling situation is casting a question on the structure of $^{26}{\rm Ne}$
PDR. Is it possible to understand the structure and decay pattern of $^{26}{\rm Ne}$ PDR
consistently?  

A possible solution for this puzzle is to explicitly include the core excitation to the
PDR. If the PDR is dominated by the neutron excitation coupled to the excited $^{25}{\rm Ne}$, the
observed decay pattern can be straightforwardly understood. In particular, the coupling of the
neutron excitation with the low-lying collective modes such as rotation and vibration
\cite{Loher2016} may play an important role, because it is well known
that the neutron excitation across $N=20$ shell gap induces strong deformation of Ne isotopes in
the island of inversion \cite{Sorlin2008}. Theoretically, the microscopic description of the
rotation and vibration coupling requires the treatment beyond the linear response. For this
purpose, we use antisymmetrized molecular dynamics (AMD) 
\cite{Kanada-Enyo2003,Kanada-Enyo2012} and its extended version
called shifted-basis AMD \cite{Kanada-Enyo2005,Chiba2015a,Kanada-Enyo2016,Kanada-Enyo2016a}. 
In this framework, by the angular momentum projection, the 
rotational motion is properly described. And, by introducing the basis wave functions in which the
centroids of the Gaussian wave packets describing nucleons are ``shifted'', it is able to describe
various particle-hole configurations. This framework has been applied to the isoscalar  monopole
and dipole responses of light stable nuclei \cite{Kanada-Enyo2014,Chiba2015a,Kanada-Enyo2016a} and
electric and isoscalar dipole responses of  neutron-rich Be isotopes
\cite{Kanada-Enyo2005,Kanada-Enyo2016}.  

In this study, the shifted-basis AMD is applied to the electric dipole response of 
$^{26}{\rm Ne}$. It is shown that the observed energy and strength of $^{26}{\rm Ne}$ PDR is
successfully described by shifted-basis AMD. Furthermore, it is found that the PDR is dominated by
the neutron excitation coupled to the quadrupole excitation of the core, which qualitatively
explains the observed decay pattern of PDR. It is also discussed that the PDR has large isoscalar 
component at the same time, because of the core excitation. 

This paper is organized as follows. The theoretical framework of shifted-basis AMD is explained in
Sec. \ref{sec:2}, and the numerical results for the low-lying spectrum of $^{24,25,26}{\rm Ne}$
and the electric dipole response of $^{26}{\rm Ne}$ are presented in Sec. \ref{sec:3}. The
analysis of the numerical results are discussed in Sec. \ref{sec:4}. We first discuss the
splitting of GDR. Then the structure of PDR and its isoscalar component are discussed. 
The final section summarizes this study.

\section{theoretical framework}\label{sec:2}
Here, we briefly explain theoretical framework of AMD and the method to extract the
single-particle energies and orbits. Then, the generator coordinate method (GCM) and shifted-basis
AMD are introduced, which are used to describe the low-lying spectrum and the highly excited $1^-$
states of $^{26}{\rm Ne}$. Using thus-obtained GCM wave functions for the ground and $1^-$ states,
the electric dipole transition strength, response function and spectroscopic factor are calculated.

\subsection{Antisymmetrized molecular dynamics}
In the AMD framework,  we use the microscopic $A$-body Hamiltonian given as,
\begin{align}
 H = \sum_{i=1}^A t(i) + \sum_{i<j}^A v_n(ij) + \sum_{i<j}^Z v_C(ij)  - t_{cm}.
\end{align}
In this study, we employ the Gogny D1S interaction \cite{Berger1991} as an effective
nucleon-nucleon interaction $v_n$ and the Coulomb interaction $v_C$ is approximated by a sum of
seven Gaussians. The center-of-mass kinetic energy $t_{cm}$ is exactly removed, which is
essentially important to remove the spurious modes from the isoscalar dipole response.  

The intrinsic wave function $\Phi_{int}$  is represented by a Slater determinant of single
particle wave packets. It is projected to the eigenstate of parity before the variation
(parity projection before variation),
\begin{align}
 \Phi_{int}&={\mathcal A} \{\varphi_1,\varphi_2,...,\varphi_A \},\\
 \Phi^\pi_{int} &= \frac{1+\pi \hat{P}_x}{2}\Phi_{int}, \quad \pi=\pm.
  \label{EQ_INTRINSIC_WF}  
\end{align}
Here $\varphi_i$ is the single nucleon wave packet having deformed Gaussian form
\cite{Kimura2004a}, 
\begin{align}
 \varphi_i({\bf r}) &= \prod_{\sigma=x,y,z}
 \left(\frac{2\nu_\sigma}{\pi}\right)^{\frac{1}{4}}
 e^{-\nu_\sigma\left(r_\sigma - 
  \frac{Z_{i\sigma}}{\sqrt{\nu_\sigma}}\right)^2+\frac{1}{2}Z^2_\sigma}\chi_i\xi_i,
 \label{eq:singlewf} 
\end{align}
where $\chi_i$ is the spinor and $\xi_i$ is the isospin fixed to proton or neutron.
The  ${\bm Z}_i$, $\bm \nu$ and $\chi_i$  are the parameters of the wave function and
determined by the energy variation which minimizes the expectation value of the Hamiltonian,
\begin{align}
 \widetilde{E}&=\frac{\langle \Phi^\pi|\hat H|\Phi^\pi\rangle}{\langle
  \Phi^\pi|\Phi^\pi\rangle} +  v_\beta(\langle\beta\rangle-\beta)^2.
\end{align}
Here the potential $v_\beta(\langle\beta\rangle-\beta)^2$ imposes the constraint on
the quadrupole deformation parameter $\braket{\beta}$ defined  in Ref. \cite{Kimura2012}.
The magnitude of $v_\beta$ is chosen large enough so that $\braket{\beta}$ equals to $\beta$ after
the energy variation. No constraint was imposed on another quadrupole deformation parameter
$\braket{\gamma}$, and hence, it always has the optimal value for each $\beta$.  As a
result of the energy variation, we obtain the optimized wave function denoted by
$\Phi^\pi_{int}(\beta)$ for each given value of $\beta$.   

\subsection{single particle levels}
To investigate the single-particle configuration of the optimized wave functions
$\Phi^\pi_{int}(\beta)$, we construct the single-particle Hamiltonian from $\Phi_{int}(\beta_i)$,
and calculate the neutron single-particle energies and orbits by diagonalizing it.  We first
transform the single particle wave packets to the orthonormalized basis, 
\begin{align}
 \widetilde{\varphi}_p(\bm r) =
 \frac{1}{\sqrt{\lambda_p}}\sum_{i=1}^{A}c_{ip}\varphi_i(\bm r).   
\end{align}
Here, $\lambda_p$ and $c_{ip}$ are the eigenvalues and eigenvectors of the
overlap matrix $B_{ij}=\langle\varphi_i|\varphi_j\rangle$. Using this basis, the single-particle
Hamiltonian is constructed,
\begin{align}
 h_{pq} &= 
  \langle\widetilde{\varphi}_p|t|\widetilde{\varphi}_q\rangle + 
  \sum_{r=1}^{A}\langle
  \widetilde{\varphi}_p\widetilde{\varphi}_r|
  {v_n+v_C}|
  \widetilde{\varphi}_q\widetilde{\varphi}_r -
 \widetilde{\varphi}_r\widetilde{\varphi}_q\rangle,\nonumber\\ 
 &+\frac{1}{2}\sum_{r,s=1}^{A}
 \langle\widetilde{\varphi}_r\widetilde{\varphi}_s
|\widetilde{\varphi}_p^*\widetilde{\varphi}_q
\frac{\delta v_n}{\delta \rho}|\widetilde{\varphi}_r
\widetilde{\varphi}_s - \widetilde{\varphi}_s  \widetilde{\varphi}_r
\rangle.
\end{align}
The eigenvectors $f_{q\alpha}$ of $h_{pq}$ defines the occupied single particle orbits
${\phi}_\alpha=\sum_{q=1}^{A}f_{q\alpha}\widetilde{\varphi}_q$ and  their eigenvalues
$\varepsilon_\alpha$ are the single-particle energies. To understand the properties of the single
particle orbits, we also calculate the amount of the positive-parity component,  
\begin{align}
 p^+ = |\langle {\phi}_s|\frac{1+P_x}{2}| {\phi}_s\rangle|^2, \label{eq:sp1}
\end{align}
and angular momenta in the intrinsic frame,
\begin{align}
 j(j+1)&= \langle {\phi}_s|{j}^2| {\phi}_s\rangle, \quad
 \Omega = \sqrt{\langle {\phi}_s|{j}_z^2| {\phi}_s\rangle},\label{eq:sp2}\\
 l(l+1)&= \langle {\phi}_s|{l}^2| {\phi}_s\rangle, \quad
 m_l = \sqrt{\langle {\phi}_s|{l}_z^2| {\phi}_s\rangle},\label{eq:sp3}
\end{align}
which corresponds to the asymptotic quantum number of the Nilsson orbits.

\subsection{Generator coordinate method\\ and shifted-basis AMD} 
To describe the ground and excited states, we perform the angular momentum projection and
GCM. We also explain the shifted-basis AMD 
\cite{Kanada-Enyo2005,Chiba2015a,Kanada-Enyo2016,Kanada-Enyo2016a}
 which is used to generate additional basis
wave functions for GCM. First, the eigenstate of the total angular momentum $J$ is projected out 
from the optimized wave functions $\Phi^{\pi}_{int}(\beta)$,
\begin{eqnarray}
 \Phi^{J^\pi M}_{K}(\beta) = \frac{2J+1}{8\pi^2}\int d\Omega D^{J*}_{MK}(\Omega)
  R(\Omega)\Phi^{\pi}_{int}(\beta). \label{eq:prjwf}
\end{eqnarray} 
Here, $D^{J}_{MK}(\Omega)$ is the Wigner $D$ function and ${R}(\Omega)$ is the rotation
operator. The integrals over  three Euler angles $\Omega$ are evaluated numerically. This
projected wave function $\Phi^{J^\pi M}_{K}(\beta)$ is used as the basis wave functions of 
GCM. 

Then, the wave functions having different quadrupole deformation $\beta$ and projection of angular
momentum $K$ are superposed (GCM),  
\begin{align}
 \Psi^{J^\pi M}_{n} = \sum_{K=-J}^J\sum_{i=1}^N
 e_{Kin}\Phi^{J^\pi M}_{K}(\beta_i).\label{eq:gcmwf0} 
\end{align}
where $N$ is a number of basis wave functions prepared by the energy variation.
The coefficients $e_{Kin}$ and eigenenergies $E^{J^\pi}_n$ are obtained by solving the 
Hill-Wheeler equation \cite{Hill1953,Griffin1957}, 
\begin{align}
 &\sum_{K'i'}{H^{J^\pi}_{KiK'i'}e_{K'i'n}} = E^{J^\pi}_n
 \sum_{K'i'}{N^{J^\pi}_{KiK'i'}e_{K'i'n}},\\  
 &H^{J^\pi}_{KiK'i'} 
 = \langle{\Phi^{J^\pi M}_{K}(\beta_i)|H|\Phi^{J^\pi M}_{K'}(\beta_{i'})}\rangle, \\
 &N^{J^\pi}_{KiK'i'} 
 = \langle{\Phi^{J^\pi M}_{K}(\beta_i)|\Phi^{J^\pi M}_{K'}(\beta_{i'})}\rangle.
\end{align}
As explained in the next section, the basis wave functions $\Phi^{J^\pi M}_{K}(\beta_i)$ are not
sufficient to describe GDR, because many of the $1p1h$ configurations which coherently contribute
to GDR are missing. To introduce various $1p1h$ configurations, we use the shifted-basis 
AMD which generates additional basis wave functions as explained below. We denote
by $X^i$ a set of parameters of the optimized wave function $\Phi^\pi_{int}(\beta_i)$,  
\begin{align}
 X^i=\set{\bm Z_1,...,\bm Z_A,\bm \nu,\chi_1,...,\chi_A}.\label{eq:set0}
\end{align}
and introduce new sets of parameters,
\begin{align}
 &X_j^i=\set{\bm Z_1',...,\bm Z_{j}',...,\bm Z'_A,\bm \nu,\chi_1,...,\chi_A},
 \nonumber\\
 &\bar{X}_j^i=\set{\bm Z_1',...,\bm Z_{j}',...,\bm Z'_A,\bm \nu,\chi_1,...,\bar{\chi}_j,...,\chi_A}.
 \nonumber\\
 &j=1,...,A. \label{eq:set1}
\end{align}
where $\bar{\chi}_j$ is the time reversal of $\chi_j$, and $\bm Z'$ is generated by shifting the
original position of the $j$th Gaussian centroid by $\epsilon \bm e_\sigma$,
\begin{align}
 &\bm Z_p' = \left\{
 \begin{array}{ll}
  (\bm Z_p + \epsilon\bm e_\sigma) -\epsilon\bm e_\sigma/A, & p=j\\
  \bm Z_p - \epsilon\bm e_\sigma/A, & p\neq j
 \end{array}
 \right.,\label{eq:shift1}\\
 &\sigma=x,y,z.
\end{align}
Here, $\bm e_\sigma$ are the unit vectors in $x$, $y$ and $z$ directions, and $\epsilon$
represents the magnitude of the shift which is typically chosen as $\epsilon=0.3$ fm in this
study. All Gaussian centroids are simultaneously shifted by  $-\epsilon\bm e_\sigma/A$ to satisfy
the relation $\sum_{p=1}^A\bm Z'_p=\sum_{p=1}^A \bm Z_p = 0$, which is needed to avoid the
contamination of the spurious center-of-mass excitation. Those new parameter sets generate
$6NA(2J+1)$ new wave functions denoted by $\Phi^{J^\pi M}_{K}(\beta_i; X_j^i)$ and 
$\Phi^{J^\pi M}_{K}(\beta_i; \bar{X}_j^i)$ to  be used as additional basis wave functions. 
The meaning of the shift of Gaussian centroids is explained in appendix \ref{sec:appb}.
If we perform GCM with only those new basis functions, the GCM wave function is given as,  
 \begin{align}
 \Psi^{J^\pi M}_{n} &= \sum_{Kij}\left(f_{Kijn}\Phi^{J^\pi M}_{K}(\beta_i;X_j)
 +g_{Kijn}\Phi^{J^\pi M}_{K}(\beta_i;\bar{X}_j)\right),\label{eq:gcmwf1}
\end{align}
and if we include all basis functions, 
\begin{align}
 \Psi^{J^\pi M}_{n} &= \sum_{Ki}e_{Kin}\Phi^{J^\pi M}_{K}(\beta_i)\nonumber\\
 +&\sum_{Kij}\left(f_{Kijn}\Phi^{J^\pi M}_{K}(\beta_i;X_j)
 +g_{Kijn}\Phi^{J^\pi M}_{K}(\beta_i;\bar{X}_j)\right),\label{eq:gcmwf2}
\end{align}
where the coefficients of superposition are determined by solving Hill-Wheeler equation.
Hereafter, we denote the GCM calculations using the wave function Eq. (\ref{eq:gcmwf0}),
(\ref{eq:gcmwf1}) and (\ref{eq:gcmwf2}) as $\beta$ GCM, shifted-basis GCM and full GCM,
respectively. 

\subsection{Dipole transition strength}
Using the GCM wave functions for the ground and excited $1^-$ states, we calculate the
electric dipole transition probability $B(E1)$ and excitation function $S(E1;E)$ defined as, 
\begin{align}
 &\mathcal{M_\mu}(E1)=\frac{N}{A}\sum_{i\in \rm p} r_iY_{1\mu}(\hat r)
 -\frac{Z}{A}\sum_{i\in \rm n} r_iY_{1\mu}(\hat r),\\
 &B(E1;0^+_1\rightarrow 1^-_n) = \sum_\mu|\braket{\Psi_{n}^{1^- \mu}|
 \mathcal{M}_\mu(E1)|\Psi_{1}^{0^+0}}|^2,\\
% &m_1(E1;\varepsilon_1,\varepsilon_2) = \sum_n \int^{\varepsilon_2}_{\varepsilon_1} dE
% \delta(E-E_n^{1^-})\nonumber\\
% &\hspace{2.5cm}\times(E - E_1^{0^+}) B(E1;0^+_1\rightarrow 1^-_n) ,\\
 &S(E1;E) = \sum_n \frac{\Gamma/2}{\pi}
 \frac{B(E1;0^+_1\rightarrow 1^-_n)}{(E-E_n)^2+\Gamma^2/4},
\end{align}
where the smearing width is chosen as $\Gamma=1$ MeV. The energy weighted and non-weighted sums, 
\begin{align}
 m_{n} = \int dE\ B(E1;E)E^n,
\end{align}
are also evaluated to see the centroid energy of GDR and the convergence of the calculation.
In addition to $E1$ response, we also calculated the isoscalar dipole transitions whose operator
is defined as 
\begin{align}
 &\mathcal{M}_\mu(IS1) = \sum_{i=1}^A (\bm r_i - \bm r_{cm})^2
 \mathcal Y_{1\mu}(\bm r_i - \bm r_{cm}),
\end{align}
where $\bm r_{cm}$ denotes the center-of-mass of the system and the solid spherical harmonics is
defined as $\mathcal Y_{lm}(\bm r)=r^lY_{lm}(\hat r)$. The transition probability $B(IS1)$ and
excitation function $S(IS1;E)$ are defined in a same manner to the $E1$ transition.

\subsection{Overlap amplitude and spectroscopic factor}
To investigate the structure of the $1^-$ states, we calculated the overlap amplitude and
spectroscopic factor. The overlap amplitude is defined as the overlap between the wave functions
of nuclei with mass $A$ and $A+1$. For example, the overlap amplitude for $^{26}{\rm Ne}$ is
defined as,
\begin{align}
 \varphi(\bm r) = \sqrt{A+1}\braket{\Psi^{J'^{\pi'} M'}_{n'}(^{25}{\rm Ne})|
 \Psi^{J^{\pi} M}_n(^{26}{\rm Ne})}. \label{eq:sfac1}
\end{align}
If the wave functions for  $^{25}{\rm Ne}$ and $^{26}{\rm Ne}$ are given by $\beta$ GCM, 
Eq. (\ref{eq:sfac1}) reads,
\begin{align}
 \varphi(\bm r) =& \sqrt{A+1}\sum_{KiK'i'}e^*_{K'i'n'}e_{Kin}\nonumber\\
 &\times\braket{\Phi_{K'}^{J'^{\pi'} M'}(\beta_{i'};{}^{25}{\rm Ne})|
 \Phi_K^{J^\pi M}(\beta_{i};{}^{26}{\rm Ne})}.
\end{align}
Using Eq. (\ref{eq:appa5}) and (\ref{eq:appa6}), it is calculated as
\begin{align}
 \varphi(\bm r) = \sum_{jl}C^{JM}_{J'M',jM-M'}\varphi_{jl}(r)
 [Y_l(\hat{r})\otimes \chi]_{jM-M'}
\end{align}
\begin{widetext}
 \begin{align}
  \varphi_{jl}(r) = \sum_{KiK'i'}e^*_{K'i'n'}e_{Kin}
  \sum_k C^{JK}_{J'K'-k,jk}\sum_{p=1}^{26}(-)^p\psi^{(p)}_{jlk}(r;i)
  \frac{2J'+1}{8\pi^2}\int d\Omega D^{J'*}_{K'K-k}(\Omega) \det B^{(p;ii')}(\Omega).
 \end{align}
\end{widetext}
Once the overlap amplitude is calculated, its integral yields the spectroscopic factor,
\begin{align}
 S_{jl} = \int_0^\infty r^2dr\ |\varphi_{jl}(r)|^2.
\end{align}
The details of above expressions are explained in appendix \ref{sec:appa}. It is straightforward
to derive corresponding expressions for shifted-basis GCM and full GCM wave functions.

\section{Results}\label{sec:3}
In this section, we first show the low-lying
level scheme of $^{24}{\rm Ne}$, $^{25}{\rm Ne}$ and $^{26}{\rm Ne}$ obtained by $\beta$ GCM. Then
we compare the electric dipole response functions obtained by $\beta$ GCM, shifted-basis GCM and
full GCM. 

\subsection{Results of energy variation and single-particle configurations}
\begin{figure*}[htb]
 \includegraphics[width=\hsize]{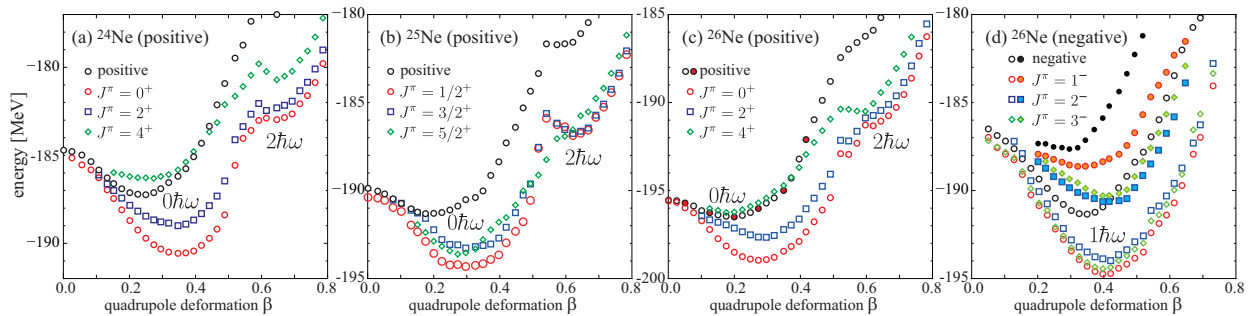}
 \caption{(color online) The energy curves as functions of quadrupole deformation parameter
 $\beta$ for $^{24}{\rm Ne}$, $^{25}{\rm Ne}$ and $^{26}{\rm Ne}$. Symbols denoted by ``positive''
 or  ``negative'' show the results of the energy variation after the parity
 projection, while others show those after the angular momentum  projection.
 For the negative-parity states of $^{26}{\rm Ne}$, two different single-particle configurations
 were obtained, which are shown by open and filled symbols.}\label{fig:curve} 
\end{figure*}
\begin{table}[h]
\begin{center}
\caption{The valence four proton and six neutron orbits of $^{26}{\rm Ne}$ at the energy minima of
 (a) positive parity with $0\hbar\omega$ configuration, (b) negative parity with  neutron
 excitation and (c) negative parity with proton excitation. The single particle energy
 $\varepsilon$ is given in MeV. Other quantities are defined by
 Eqs. (\ref{eq:sp1})-(\ref{eq:sp3}).  The Nilsson asymptotic quantum numbers
 $[Nn_zm_l\Omega^\pi]$ deduced from those  properties are also given.} 
 \label{tab:spo} 
 \begin{ruledtabular}
  \begin{tabular}{cccccccc} 
   \multicolumn{8}{c}{(a) positive minimum at $\beta=0.30$}\\\hline
   orbit & $\varepsilon $ & $p^+$ & $j$ & $l$ & $m_{l}$ & $\Omega$ &$[Nn_zm_l\ \Omega^\pi]$
   \\\hline 
   $\pi$ 1, 2 & $-17.0$ & 0.99 & 2.3 & 1.9 & 0.5 & 0.5 & $[220\ 1/2^+]$\\
   $\pi$ 3, 4 & $-24.3$ & 0.01 & 0.7 & 1.0 & 1.0 & 0.5 & $[101\ 1/2^-]$\\ \hline
%   neutron &$\varepsilon $ & $p^+$ & $j$ & $l$ & $m_{l}$ & $\Omega$ &$[Nn_zm_l\ \Omega^\pi]$
%   \\ \hline
   $\nu$ 1, 2 & $-5.2 $ & 0.99 & 1.5 & 1.5 & 0.8 & 0.5 & $[211\ 1/2^+]$\\
   $\nu$ 3, 4 & $-7.3 $ & 0.99 & 2.5 & 2.0 & 2.0 & 2.5 & $[202\ 5/2^+]$\\ 
   $\nu$ 5, 6 & $-10.1$ & 0.99 & 2.5 & 2.0 & 1.1 & 1.5 & $[211\ 3/2^+]$\\\\
   \multicolumn{8}{c}{(b) negative minimum at $\beta=0.40$ (neutron ex.)}\\\hline
   orbit & $\varepsilon $ & $p^+$ & $j$ & $l$ & $m_{l}$ & $\Omega$  & $[Nn_zm_l\ \Omega^\pi]$
   \\ \hline
   $\pi$ 1, 2 & $-18.8$ & 0.99 & 2.3 & 1.8 & 0.4 & 0.5 & $[220\ 1/2^+]$\\
   $\pi$ 3, 4 & $-23.1$ & 0.10 & 0.7 & 1.1 & 1.0 & 0.5 & $[101\ 1/2^-]$\\ \hline
%   neutron & $\varepsilon $ & $p^+$ & $j$ & $l$ & $m_{l}$ & $\Omega$ &$[Nn_zm_l\ \Omega^\pi]$
%   \\ \hline
   $\nu$ 1   & $-1.4 $ & 0.00 & 2.9 & 2.5 & 0.4 & 0.5 & $[330\ 1/2^-]$\\
   $\nu$ 2   & $-5.2 $ & 0.99 & 1.7 & 1.7 & 0.9 & 0.5 & $[211\ 1/2^+]$\\ 
   $\nu$ 3, 4 & $-6.3 $ & 0.99 & 2.5 & 2.0 & 2.0 & 2.5 & $[202\ 5/2^+]$\\ 
   $\nu$ 5, 6 & $-10.1$ & 0.99 & 2.5 & 2.1 & 1.1 & 1.5 & $[211\ 3/2^+]$\\ \\
   \multicolumn{8}{c}{(c) negative minimum at $\beta=0.32$ (proton ex.)}\\\hline
   orbit & $\varepsilon $ & $p^+$ & $j$ & $l$ & $m_{l}$ & $\Omega$  &$[Nn_zm_l\ \Omega^\pi]$
   \\ \hline
   $\pi$ 1, 2 & $-16.7$ & 0.99 & 2.4 & 1.9 & 0.6 & 0.7 & $[220\ 1/2^+]$\\
   $\pi$ 3   & $-20.1$ & 0.53 & 1.8 & 1.6 & 1.0 & 1.0 & $[211\ 3/2^+]+ [101\ 1/2^-]$\\ 
   $\pi$ 4   & $-25.0$ & 0.00 & 0.7 & 1.0 & 1.0 & 0.5 & $[101\ 1/2^-]$\\ \hline
%   neutron & $\varepsilon $ & $p^+$ & $j$& $l$ & $m_{l}$ & $\Omega$ &$[Nn_zm_l\ \Omega^\pi]$
%    \\ \hline
   $\nu$ 1, 2 & $-5.1$  & 0.99 & 1.5 & 1.4 & 0.8 & 0.6 & $[211\ 1/2^+]$\\
   $\nu$ 3, 4 & $-7.2$  & 0.99 & 2.5 & 2.0 & 2.0 & 2.5 & $[202\ 5/2^+]$\\ 
   $\nu$ 5, 6 & $-10.3$ & 0.99 & 2.5 & 2.0 & 1.1 & 1.5 & $[211\ 3/2^+]$\\
  \end{tabular}
  \end{ruledtabular}
\end{center}
\end{table}

\begin{figure}[h]
 \includegraphics[width=1.0\hsize]{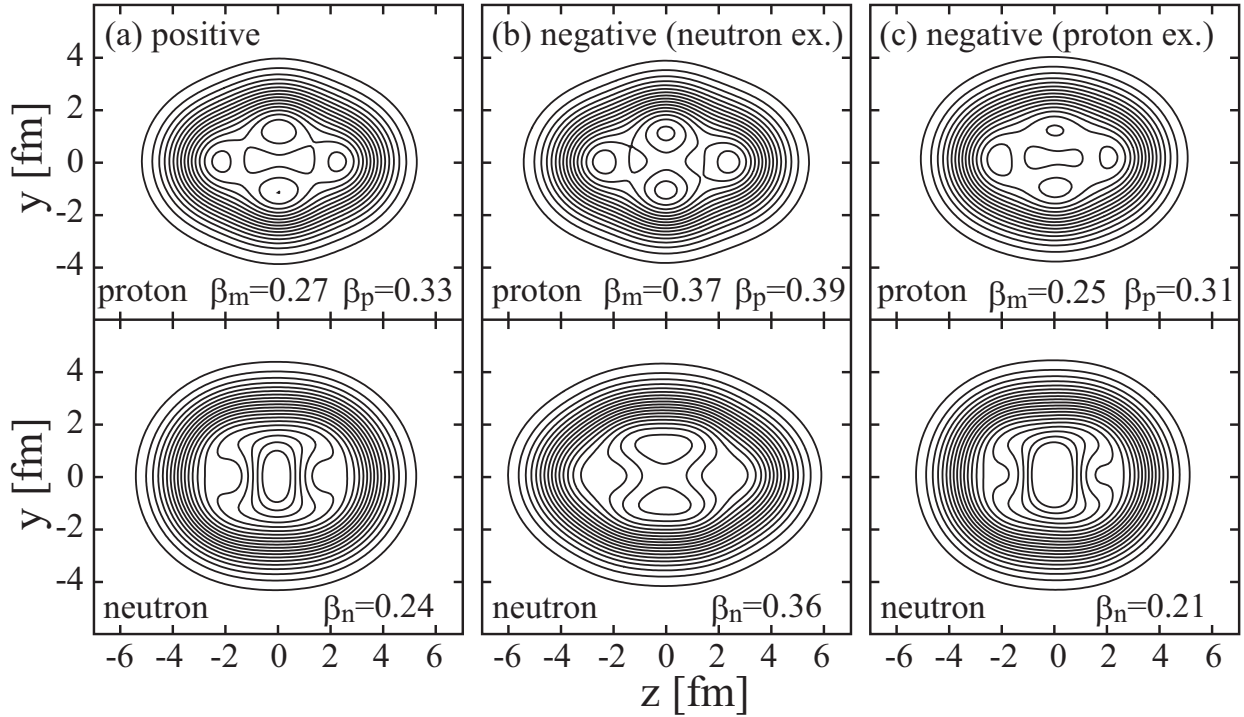}
 \caption{Intrinsic density distributions of $^{26}{\rm Ne}$ at energy minima of positive- and
 negative-parity states. Upper (lower) panels show proton (neutron)
 distributions.}\label{fig:dens}   
\end{figure}

Figure \ref{fig:curve} shows the energy curves for positive-parity states of $^{24}{\rm Ne}$ and
$^{25}{\rm Ne}$, and those for positive- and negative-parity states of $^{26}{\rm Ne}$ obtained by the
energy variation after the parity projection and the angular momentum projection. All nuclei
discussed here locate out of the island of inversion, and hence, their ground states are dominated
by the $0\hbar\omega$ (normal) configurations.  The  strongly deformed $2\hbar\omega$ (intruder)
configurations  locate approximately 7 MeV above the normal configurations in all nuclei. After
the angular momentum projection, the energy minima of $0^+$ or $1/2^+$ states corresponding to the
ground states  have non-negligible deformations that are $\beta\simeq 0.35$ for $^{24}{\rm Ne}$ and
$\beta\simeq 0.30$ for  $^{25}{\rm Ne}$  and  $^{26}{\rm Ne}$. For the negative-parity states of
$^{26}{\rm Ne}$, we have obtained two energy minima which have different internal structures.  

The single-particle configurations of the positive- and negative-parity minima of $^{26}{\rm Ne}$
can be understood from the properties of single-particle  orbits listed in Tab. \ref{tab:spo}. At
the energy minimum of the positive-parity state, the most weakly bound two neutrons occupy the
$[211\ 1/2^+]$ Nilsson orbit which originates in the spherical $1s_{1/2}$ orbit
(Tab. \ref{tab:spo} (a)). Owing to the spherical nature of this orbit, the deformation of neutron
distribution is smaller than that of proton distribution as seen its density profile shown in
Fig. \ref{fig:dens} (a), which reduces the deformation of the system compared to $^{24}{\rm Ne}$
as mentioned above.  

The $1^-$ states shown by open circles in Fig. \ref{fig:curve} (d) have the energy minimum located
around $\beta=0.40$ at approximately 4 MeV above the positive-parity minimum, whose
single-particle levels are listed in Tab. \ref{tab:spo} (b). We see that the  protons do not
change their configuration from the positive-parity minimum,  but a valence neutron is excited
from the $[211\ 1/2^+]$ orbit to the $[330\ 1/2^-]$ orbit  which corresponds to the neutron
excitation from $sd$- to $pf$-shells. Those neutron particle and hole enlarge the deformation of
the neutron distribution, and as a result, the deformation of the system is much larger than the
positive-parity minimum (Fig. \ref{fig:dens} (b)). It is noted that the degeneracy of the single
particle orbit is lost in this configuration because the time reversal symmetry is
broken. Therefore, the single particle energies and  other properties listed in the table are
averaged for the pair of the approximately  degenerated orbits. Another $1^-$ states shown by
filled circles in Fig. \ref{fig:curve} (c) have the minimum approximately 8 MeV above the
positive-parity minimum around $\beta=0.32$ whose single-particle configuration is given in
Tab. \ref{tab:spo} (c). In this state, the neutron configuration is unchanged from the
positive-parity minimum, but the third proton occupies the orbit which is an admixture of the
positive- and negative-parity.  From the properties of this orbit, we deduced that  the 
$[211\ 3/2^+]$ and $[101\ 1/2^-]$ orbits are mixed. Therefore, when the intrinsic wave function is
projected to the negative  parity, this configuration approximately corresponds to the proton
$[211\ 3/2^+]$ particle and  $[101\ 1/2^-]$ hole state. Those proton particle and hole reduce the
deformation of the proton distribution (Fig. \ref{fig:dens} (c)) leading to the reduction of the
total system deformation.  The $J^\pi=2^-$ and $3^-$ states shown by open and filled boxes and
diamonds have the same single particle configurations with $J^\pi=1^-$ states mentioned above.  

Although we do not show the calculated results, it is noted that the single-particle configurations
of $^{\rm 24}{\rm Ne}$ and $^{25}{\rm Ne}$ are understood in the same way. Namely, their ground
states are dominated by the normal configuration, while the  2$\hbar \omega$ excited configuration
has two neutrons in the $[330\ 1/2^+]$ orbit. The negative-parity states of $^{25}{\rm Ne}$ also
have two energy minima. The lowest minimum has a neutron excitation from the $[211\ 1/2^+]$ orbit
to the $[330\ 1/2^+]$ orbit, while the upper minimum has a proton excitation from the 
$[101\ 1/2^-]$ orbit to the $[211\ 3/2^+]$ orbit, which is qualitatively same with $^{26}${\rm Ne}.   

The results obtained by the energy variation and angular momentum projection are summarized as
follows. (1) $^{25}{\rm Ne}$ and $^{26}{\rm Ne}$ have positive-parity  minimum with smaller
deformation compared to that of $^{24}{\rm Ne}$. This is because of the valence neutrons occupying
the $[211\ 1/2^+]$ orbit which originates in the spherical $1s_{1/2}$ orbit. (2) $^{25}{\rm Ne}$
and $^{26}{\rm Ne}$ has two negative-parity minima having different single particle
configurations. The lowest minimum  has a neutron $[330\ 1/2^-]$ particle 
and a $[211\ 1/2^+]$  hole. Those neutron particle-hole enlarge the deformation compared to the
positive-parity minimum. (3) Another minimum  has a proton $[211\ 3/2^+]$ particle and 
$[101\ 1/2^-]$ hole, which reduces the nuclear deformation.

\subsection{Low-lying energy spectra obtained by $\bm \beta$ GCM} 
\begin{figure*}[htb!]
 \includegraphics[width=0.9\hsize]{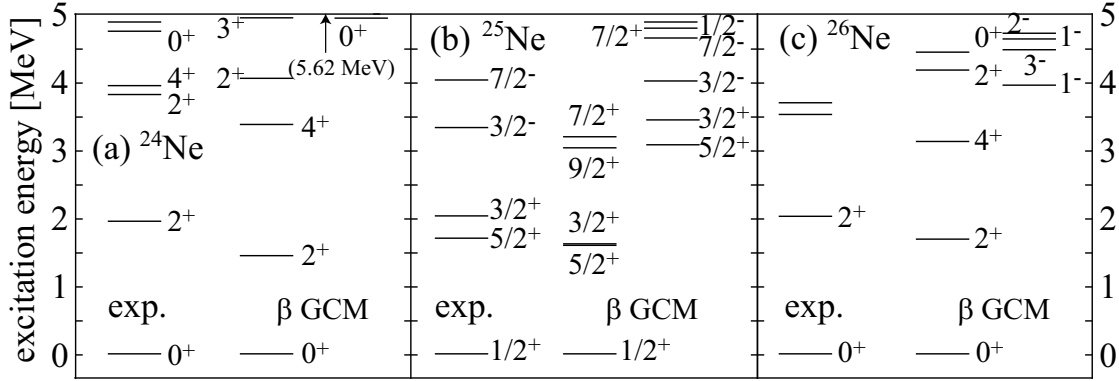}
 \caption{Observed \cite{Padgett2005,Belleguic2005,Obertelli2006,Catford2010} and calculated
 spectrum of $^{24}{\rm Ne}$,  $^{25}{\rm Ne}$ and  $^{26}{\rm Ne}$. }\label{fig:spectra} 
\end{figure*}

The energy spectra of $^{24}{\rm Ne}$, $^{25}{\rm Ne}$ and $^{26}{\rm Ne}$ obtained by $\beta$ GCM
are shown in Fig. \ref{fig:spectra}. We first examine the result for $^{24}{\rm Ne}$. The ground
band members ($0^+$, $2^+$ and $4^+$ states) having normal configurations are reasonably
described showing a vibrational spectrum, although the excitation energies of the first $2^+$
and  $4^+$ states are slightly underestimated. The present result also gives the reasonable
$B(E2;0^+\rightarrow 2^+)$ value which is consistent with the observed value as listed in
Tab. \ref{tab:e2}. The second $2^+_2$ state is followed by the $3^+$ state   
at approximately 5 MeV which also has normal configuration and constitute $\gamma$
vibrational band. Their relatively  small excitation energies imply the $\gamma$ softness of this
nucleus, although the $3^+$ state has not been  experimentally identified yet. The $0^+_2$ state
having the intruder configuration locates at 5.62 MeV which slightly overestimates the observed
excitation energy of 4.77 MeV. This may be due to the overestimation of the $N=20$ shell gap by
Gogny  D1S interaction which can be confirmed in the result of $^{25}{\rm Ne}$ explained below. We
have not obtained any negative-parity states below 5 MeV in this nucleus.   

\begin{table}[h]
\begin{center}
\caption{Reduced $E2$ transition probabilities for low-lying states of neon isotopes in the unit
 of $e^2{\rm fm}^4$. Numbers in parenthesis are the experimental values taken from Refs. 
 \cite{Raman1987,Gibelin2007}}
 \label{tab:e2} 
 \begin{ruledtabular}
  \begin{tabular}{clclcl} 
   \multicolumn{2}{c}{$^{24}{\rm Ne}$}&\multicolumn{2}{c}{$^{25}{\rm Ne}$}&\multicolumn{2}{c}{$^{26}{\rm Ne}$}\\
   $J_i\rightarrow J_f$ & $B(E2)$&$J_i\rightarrow J_f$ & $B(E2)$&$J_i\rightarrow J_f$ & $B(E2)$\\
   \hline
   $2^+_1\rightarrow 0^+_1$ & 24.3 (28)&$3/2^+_1\rightarrow 1/2^+_1$ &  35.9 &$2^+_1\rightarrow 0^+_1$ & 31.1 (28)\\
   $2^+_2\rightarrow 0^+_1$ & 2.3        &$5/2^+_1\rightarrow 1/2^+_1$ &  27.4 &$2^+_2\rightarrow 0^+_1$ & 6.7 \\
   $4^+_1\rightarrow 2^+_1$ & 17.5       &$5/2^+_1\rightarrow 3/2^+_1$ &  1.6  &$4^+_1\rightarrow 2^+_1$ & 32.1 \\
   $4^+_1\rightarrow 2^+_2$ & 1.5        &$7/2^+_1\rightarrow 3/2^+_1$ &  40.9 &$4^+_1\rightarrow 2^+_2$ & 2.3 \\
   $3^+_1\rightarrow 2^+_1$ & 6.5        &$7/2^+_1\rightarrow 5/2^+_1$ &  6.5  &$3^+_1\rightarrow 2^+_1$ & 6.9 \\
   $3^+_1\rightarrow 2^+_2$ & 18.8       &$9/2^+_1\rightarrow 5/2^+_1$ &  26.1 &$3^+_1\rightarrow 2^+_2$ & 1.3 \\
                            &            &$9/2^+_1\rightarrow 7/2^+_1$ &  1.9  &   & \\
  \end{tabular}
  \end{ruledtabular}
\end{center}
\end{table}

\begin{table}[h]
\begin{center}
\caption{Single-particle spectroscopic factors for the low-lying states of 
 $^{25}{\rm  Ne}$ and the ground  state of $^{26}{\rm Ne}$. 
 Numbers in parenthesis are the experimental values taken from Refs.
 \cite{Terry2006,Catford2010}}  \label{tab:sfac} 
 \begin{ruledtabular}
  \begin{tabular}{ccccc} 
   & $0^+_1\otimes l_j$& $2^+_1\otimes s_{1/2}$
   & $2^+_1\otimes d_{5/2}$& $2^+_1\otimes d_{3/2}$\\\hline
   $^{25}{\rm Ne}(1/2^+_1)$ &0.57 (0.80)&      & 0.52 & 0.14\\
   $^{25}{\rm Ne}(3/2^+_1)$ &0.21 (0.44)& 0.15 & 0.00 & 0.05\\
   $^{25}{\rm Ne}(5/2^+_1)$ &0.07 (0.10)& 0.49 & 0.10 & 0.10\\\\
   & $0^+_1\otimes l_j$& $2^+_1\otimes p_{3/2}$
   & $2^+_1\otimes f_{7/2}$\\\hline
   $^{25}{\rm Ne}(3/2^-_1)$ &0.40 (0.75)& 0.26 & 0.08 & \\
   $^{25}{\rm Ne}(7/2^-_1)$ &0.64 (0.73)& 0.04 & 0.05 & \\\\
   & $1/2^+_1\otimes s_{1/2}$& $3/2^+_1\otimes d_{3/2}$
   & $5/2^+_1\otimes d_{5/2}$& \\\hline
   $^{26}{\rm Ne}(0^+_1)$ & 1.04 (1.4)& 0.55 (0.5)& 1.47 (1.3)
  \end{tabular}
  \end{ruledtabular}
\end{center}
\end{table}

The spectrum of $^{25}{\rm Ne}$ is shown in Fig. \ref{fig:spectra} (b). In the low-lying
positive-parity states, similar to $^{26}{\rm Ne}$, the last neutron occupies the 
$[211\ 1/2^+]$ orbit which originates in the spherical $1s_{1/2}$ orbit. Therefore, the ground
state is the $1/2^+$ state and it is followed by the $3/2^+$ to $9/2^+$ states to constitute the
ground band. The energies of the band member states ($3/2^+$ and $5/2^+$ states) reasonably agrees 
with the observation, although the $7/2^+$ and $9/2^+$ states have not been observed yet. 
%There may
%be two possible interpretation for this ground band. The first interpretation is the strong
%coupling of a valence neutron in the $[211\ 1/2^+]$ orbit to the  ground band of $^{24}{\rm Ne}$
%which result in the formation of the   $K^\pi=1/2^+$ rotational band with Coliori decoupling. The
%moment of intertia and decoupling parameter obtained by the $\chi^2$ fitting are $I=XXX$ and
%$a=XXX$, respectively. 
This ground band is interpreted as the coupling of the $1s_{1/2}$ neutron to the
ground band of  $^{24}{\rm Ne}$. Namely, $^{24}{\rm Ne}(2^+)\otimes 1s_{1/2}$ yields the 
$5/2^+$- $3/2^+$ doublet and $^{24}{\rm Ne}(4^+)\otimes 1s_{1/2}$ yields the $9/2^+$- $7/2^+$
doublet. The calculated and observed $B(E2)$ values and the spectroscopic factors listed in
Tab. \ref{tab:e2} and \ref{tab:sfac} look supporting this interpretation, although the
$0^+_1\otimes d_{3/2}$ component is not small for the $3/2^+$ state.   
%Because the 
%$B(E2\uparrow)$ strengths to the $3/2^+$ and $5/2^+$ states should be in the same magnitude in
%this interpretatoin, and the large spectroscopic factor suggest the single-particle nature of
%those states.  

The negative-parity states of $^{25}{\rm Ne}$ provides information about the size of the $N=20$
shell gap in this mass region. The calculated $3/2^-$ and $7/2^-$ states locate at 4.0 and 4.7
MeV and slightly overestimate the observed values of 3.3 and 4.0 MeV
\cite{Reed1999,Terry2004,Catford2005,Terry2006,Obertelli2006,Catford2010}. The observed and
calculated spectroscopic factors for those negative-parity states are large, and hence, their
excitation energies are good measures for the $p_{3/2}$ and $f_{7/2}$ single-particle
energies. The overestimation of the negative-parity state energies means that the $N=20$ shell gap
given by Gogny D1S interaction is slightly larger than the experiment. The larger $N=20$ shell gap
also explains why the energy of the  $0^+_2$ state in $^{24}{\rm Ne}$ having the intruder
configuration is also overestimated by the present calculation. It is interesting to note that the
order of the ${3/2}^-$ and ${7/2}^-$ ($1p_{3/2}$ and $0f_{7/2}$) are already inverted in this
nucleus, which  explains reason of the $p_{3/2}$ neutron-halo formation in $^{31}{\rm Ne}$
\cite{Nakamura2009,Hamamoto2010,Horiuchi2010,Minomo2012,Takechi2012}.  

The low-lying spectrum of $^{26}{\rm Ne}$ is shown in Fig. \ref{fig:spectra} (c). Experimentally,
the first $2^+$ state is known at 2.0 MeV and two states without definite spin-parity
assignment are observed at 3.5 and 3.7 MeV. The present calculation yields the first $2^+$ state
at 1.8 MeV and predict $4^+$ state at 3.2 MeV. Those yrast states constitute the ground band
dominated by the $[211\ 1/2^+]^2$ configuration or by the $(1s_{1/2})^2$ configuration, which is
confirmed from the observed and calculated spectroscopic factors listed in Tab. \ref{tab:sfac}.
The calculation also predicts the second $0^+$ state at 4.5 MeV which is dominated by the intruder
$[330\ 1/2^+]^2$ configuration. Similar to $^{24}{\rm Ne}$, this nucleus also has the low-lying
$K^\pi=2^+$ band owing to its softness against the $\gamma$ deformation. It is constituted by the
second $2^+$ state at 4.2 MeV and $3^+$ state at 5.1 MeV. Those non-yrast states may correspond
to one of the observed state at 3.5 and 3.7 MeV.  

The low-lying $1^-$ states of $^{26}{\rm Ne}$ are of particular interest because of their
relationship to the pygmy dipole resonance. The lowest energy minium in Fig. \ref{fig:curve} (c)
which is dominated by a neutron excitation yields a group of the negative parity states around 4
to 5 MeV  shown in Fig. \ref{fig:spectra} (c). It generates $1^-_1$ and $1^-_2$ states at 4.0 and
4.5 MeV. The energy minimum with a proton excitation yields another group of $1^-$ states
around 8 to 11 MeV.

\subsection{Electric dipole response of $^{\bf 26}$Ne}
\begin{figure*}[t!]
 \includegraphics[width=0.9\hsize]{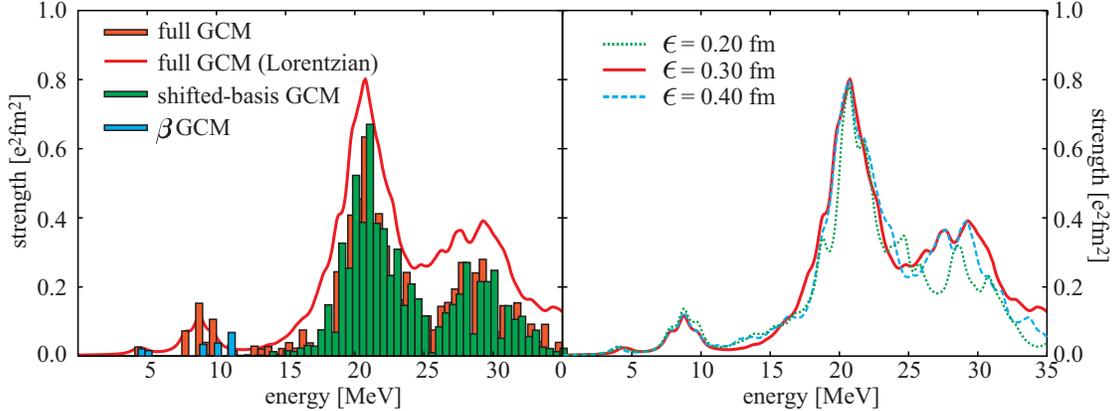}
 \caption{(a) The electric dipole strength functions of $^{26}{\rm Ne}$. The histograms show the
 results calculated by $\beta$ GCM, shifted-basis GCM  and full GCM summed in the energy bin of
 0.5 MeV width. The solid line shows the result of full GCM smeared by the Lorentzian with 1 MeV
 width. (b) Comparison of the results of full GCM calculations in which the magnitude of the shift
 is changed from 0.20 fm to 0.40 fm.}\label{fig:strength}
\end{figure*}
Figure. \ref{fig:strength} (a) shows the electric dipole strength functions where the
histograms show the results of $\beta$ GCM, shifted-basis GCM and fill GCM. The solid line
shows the full GCM result smeared with the Lorentzian with 1 MeV width. In the result of $\beta$ 
GCM (blue histogram), there are tiny peaks around 5 to 10 MeV which  are the neutron and proton
excited states explained in the previous section. On the other hand, there are almost no prominent
strength above 10  MeV, which means that $\beta$ GCM is insufficient to describe the highly
excited $1^-$ states, in particular, the GDR to which various $1p1h$ configurations coherently
contributes. 
\begin{table}[h]
\begin{center}
\caption{Energy weighted sum in $\rm e^2fm^2MeV$ and centroid energy of GDR (peak
 position and the ratio of energy weighted and non-weighted sums) in MeV obtained by $\beta$ GCM,
 shifted-basis GCM and full GCM. They are compared with the QRPA calculations
 \cite{Peru2007,Hashimoto2012} which
 also use  Gogny D1S interaction.} 
 \label{tab:sum} 
 \begin{ruledtabular}
  \begin{tabular}{cccc} 
         & $m_1$  &$m_1/m_0$ &peak pos.\\
   \hline
   $\beta$ GCM & 5 & 11 & \\
   shifted-basis GCM & 177& 24.5& 22.0\\
   full GCM & 183 & 23.4 & 21.5\\
   Peru {\it et al.} \cite{Peru2007} & & & 21.9\\
   Hashimoto \cite{Hashimoto2012} & 181 & 24.5& 22.5\\
  \end{tabular}
  \end{ruledtabular}
\end{center}
\end{table}

The shifted-basis GCM (green histogram in Fig. \ref{fig:strength} (a)) overcomes this problem. It
yields two large peaks around 21 and 28 MeV which corresponds to the GDR. The origin of this
splitting is attributed to the deformation of the ground state and discussed in the next
section. The energy weighted sum listed in Tab. \ref{tab:sum} is evidently increased compared with
$\beta$ GCM and it is consistent with other theoretical calculations with Gogny D1S interaction.
Thus, the shifted-basis GCM successfully describes GDR by introducing various $1p1h$ configurations
using the shifted Gaussian wave packets. However, the tiny peaks around 5 to 10 MeV are not clear
in the shifted-basis GCM compared to the  $\beta$ GCM. This may mean that the single particle wave
functions such as  $[330\ 1/2^-]$ and  $[101\ 1/2^-]$ that generate low-lying peaks cannot be
descried properly by the simple shift of the Gaussian basis. 

The full GCM includes all of the basis wave functions which are the single-particle excited states
obtained by the energy variation and the various $1p1h$ configurations generated by the shifted
Gaussian basis. Therefore, we expect both of the collective and single-particle excitations are
reasonably described. The strength function obtained by the full GCM is shown by orange histogram
and red line in Fig. \ref{fig:strength} (b). It has two peaked GDR distribution similar to
the shifted-GCM and low-lying strengths around 5 to 10 MeV which should be attributed to the pygmy
dipole resonance. The calculated energy weighted sum and GDR energy are similar to the result of
the shifted-basis GCM and other theoretical calculations.

Finally, we examine the convergence of the full GCM calculation. If the model space spanned by the
shifted-basis functions is large enough and if the magnitude of the shift $\epsilon$ is small
enough, the result should not depend on the magnitude of $\epsilon$. To investigate the
convergency, we performed full GCM calculations  by changing the magnitude of the $\epsilon$ to
0.2 and 0.4 fm as shown in Fig. \ref{fig:strength} (b). It is clear that the strength
distribution below 25 MeV is almost unchanged, while the peak around 28 MeV is slightly affected. 
Therefore, we conclude that the result for the pygmy dipole resonance and the first lower peak of
GDR is well converged, while the higher peak of GDR is somewhat ambiguous. We also note that the
energy weight sum of the strength and the centroid energy of GDR are rarely affected by the choice
of $\epsilon$ as shown in Tab. \ref{tab:sum}.

\section{Discussions}\label{sec:4}
Here, we first focus on the high-energy part of the calculated $E1$ response and discuss the
splitting of the GDR and its relationship to the ground state deformation. Then, we discuss the
low-energy part {\it i.e.} the PDR and analyze its characteristics.

\subsection{Splitting of GDR}
It is well known that the ground state deformation affects the distribution of the giant
resonances. In the case of the $E1$ response of axially symmetric nucleus, the ground state
deformation differentiates the oscillator length for the collective vibration along the longest
and shortest  deformation axes, which results in the splitting of the GDR into two components.
The QRPA calculation \cite{Yoshida2008a,Yoshida2009} shown that the $K^\pi=0^-$ component of the
GDR appears at smaller excitation energy than the $K^\pi=\pm 1^-$ component for the prolate
deformed nucleus. 

However, the discussion made by QRPA calculations is based on the analysis in the body-fixed frame
where the deformed intrinsic state is not an eigenstate of good angular momentum, and hence, the
calculated results do not directly correspond to the observed excitation function of $1^-$ states. 
On the other hand, in the present calculation, the results can be directly compared
with the observed data, because the rotational symmetry is restored by the angular momentum
projection. Since the excitation function shown in Fig. \ref{fig:strength} also shows the
splitting of GDR, it is of interest to check if it really originates in the ground state
deformation or not. 
\begin{figure*}[t!]
 \includegraphics[width=0.9\hsize]{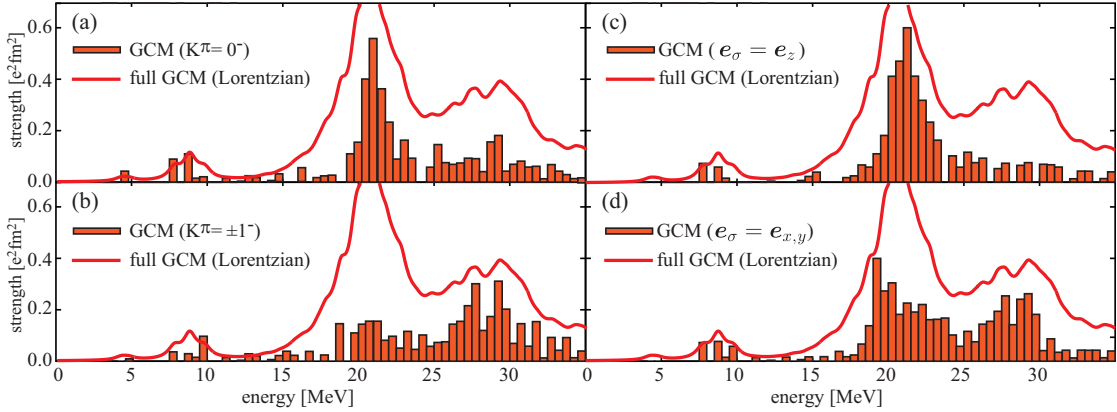}
 \caption{(a) The electric dipole strength functions of $^{26}{\rm Ne}$. The histograms show the
 results calculated by $\beta$ GCM, shifted-basis GCM  and full GCM summed in the energy bin of
 0.5 MeV width. The solid line shows the result of full GCM smeared by the Lorentzian with 1 MeV
 width. (b) Comparison of the results of full GCM calculations in which the magnitude of the shift
 is changed from 0.20 fm to 0.40 fm.}\label{fig:strength3}
\end{figure*}
For this purpose, we have performed two additional GCM calculations.
In the second calculation, the direction of the shift is unrestricted, but the value of the $K$
quantum number is restricted to $K=0$ (or $\pm 1$) in the GCM calculation. In other words, the
summation over $K$ in Eq. (\ref{eq:gcmwf2}) is restricted to only $K=0$ or $K=\pm 1$, which will
distinguish the $K=0$ and $\pm 1$ components. In the second calculation, we restricted the shift
of Gaussian centroids (the unit vector $\bm e$ in Eq. (\ref{eq:shift1})) to only the $z$ direction
(or $x$ and $y$ direction) where $z$ axis is chosen to be the longest deformation axis. This will
apparently restrict the direction of the vibration to $z$ ($x$ and $y$) direction. The results of 
the calculations are presented in Fig. \ref{fig:strength3}. As clearly seen, both calculations
show that the low-energy part of GDR is dominated by the vibration along the longest deformation
axis ($K^\pi=0^-$ and $\bm e=\bm e_z$), while the high-energy part is dominated by the  vibration
along the shortest deformation axis ($K^\pi=\pm1^-$ and $\bm e=\bm e_{x,y}$).

Thus, the splitting of the vibration modes parallel and perpendicular to the longest axis in
the intrinsic frame can be also observed even after the angular momentum projection. Hence we can
safely conclude that the splitting of GDR is surely originates in the ground state deformation.
It is also noted that the low (high) energy part of PDR is also dominated by $K^\pi=0^-$ 
($\pm 1^-$) component, which is also qualitatively consistent with the QRPA result
\cite{Yoshida2008a}. 

\subsection{Property of PDR}\label{sec:4b}
\begin{figure}[h!]
 \includegraphics[width=0.8\hsize]{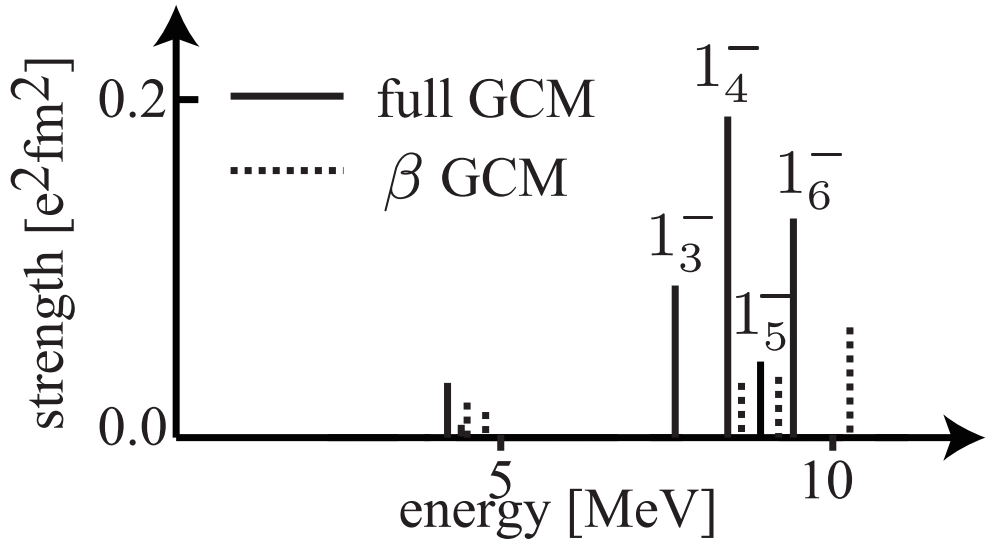}
 \caption{Distributions of the low-lying E1 strengths calculated by full GCM and $\beta$
 GCM.}\label{fig:strength2}  
\end{figure}

Figure \ref{fig:strength2} magnifies the low-energy part of the strength function.  We regard the
strength distributed from 7 to 10 MeV as PDR which consists of four $1^-$ 
states ($1^-_3$ to $1^-_6$ states) whose excitation energies and $E1$ strengths are summarized in 
Tab. \ref{tab:e1}. The averaged energy is 8.5 MeV and sum of $B(E1)$ is 0.44 $\rm e^2fm^2$, which
reasonably agrees with the observed data, 9 MeV and $0.49\pm0.16\ \rm e^2fm^2$. The EWS amounts to  
approximately 4\% of TRK sum rule, which is slightly smaller than the observation which amount to
approximately 5 \%. 
\begin{table}[h]
\begin{center}
\caption{Energies and $B(E1)$ strengths of four $1^-$ states ($1^-_3$ to $1^-_6$) which constitute
 the PDR. Their averaged energy weighted by $B(E1)$ strength and the sum of the $B(E1)$ strength
 are compared with the observed data \cite{Gibelin2008}}
 \label{tab:e1} 
 \begin{ruledtabular}
  \begin{tabular}{ccc} 
         & $E_x [\rm MeV]$  &$B(E1;0^+_1\rightarrow 1^-)\ [\rm e^2fm^2]$ \\
   \hline
   $1^-_3$  & 7.6 & 0.09  \\
   $1^-_4$  & 8.4 & 0.19  \\
   $1^-_5$  & 8.9 & 0.04  \\
   $1^-_6$  & 9.4 & 0.12  \\
   \hline
   Total    & 8.5 & 0.44  \\
   Exp.     & 9 & 0.49$\pm 0.16$  \\
  \end{tabular}
  \end{ruledtabular}
\end{center}
\end{table}
\begin{table}[h]
\begin{center}
\caption{Single-particle spectroscopic factors in $^{25}{\rm Ne}(J^\pi)\otimes \ell_j$ channels
 for the low-lying states of the $1^-$ states which  constitute the PDR.} 
 \label{tab:sfac2} 
 \begin{ruledtabular}
  \begin{tabular}{ccccc} 
   & $1/2^+_1\otimes p_{3/2}$& $1/2^+_1\otimes p_{1/2}$
   & $3/2^+_1\otimes p_{3/2}$& $3/2^+_1\otimes p_{1/2}$\\\hline
   $1^-_3$ & 0.1 & 0.1 & 0.4 & 0.0\\
   $1^-_4$ & 0.3 & 0.0 & 0.3 & 0.1\\
   $1^-_5$ & 0.2 & 0.0 & 0.2 & 0.1\\
   $1^-_6$ & 0.1 & 0.1 & 1.1 & 0.2\\\\
   & $5/2^+_1\otimes p_{3/2}$& $5/2^+_1\otimes f_{7/2}$
   & $3/2^-_1\otimes s_{1/2}$& $3/2^-_1\otimes d_{3/2}$\\\hline
   $1^-_3$ & 1.2  & 0.9 & 0.4 & 0.2 \\
   $1^-_4$ & 1.1  & 0.6 & 0.3 & 0.1\\
   $1^-_5$ & 0.3  & 0.1 & 0.7 & 0.4\\
   $1^-_6$ & 0.2  & 0.1 & 0.5 & 0.3\\
  \end{tabular}
  \end{ruledtabular}
\end{center}
\end{table}

Table \ref{tab:sfac2} suggests that there are several interesting features to be noted in the
calculated S-factors of the PDR. Firstly, the S-factors for the ground state of
$^{25}{\rm Ne}$ ($^{25}{\rm Ne}(1/2^+_1)$) is smaller than those for the excited states  
($^{25}{\rm Ne}(5/2^+_1)$, $^{25}{\rm Ne}(3/2^+_1)$ and $^{25}{\rm Ne}(3/2^-_1)$) indicating that
the PDR of $^{26}{\rm Ne}$ involves the core excitation. This may be a straightforward answer to
the question, ``Why the observed $^{26}{\rm Ne}$ PDR predominantly decays to the excited state of  
$^{25}{\rm Ne}$, not to its ground state?''. The reason of the core excitation may be attributed to 
the deformation of PDR. In the strong coupling picture, it can be easily shown that PDR has large
amount of the core excited component. Another possible reason is the isoscalar component in
PDR. I'll discuss, in the next section, that the large IS component in PDR possibly induces strong
quadrupole core excitation. 

The second is the dominance of the $p_{3/2}$ S-factors over the $f_{7/2}$ S-factors. There may be
several explanations for this. First reason is derived from a simple spherical shell model
picture. In the spherical shell model, the last neutron occupies $1s_{1/2}$ in the ground
state. This last neutron must be excited to $p_{3/2}$ not to $f_{7/2}$ to generate $1^-$
state. The second is given by the deformation picture. As already shown, the PDR has large amount
of the $\nu[330\ 1/2^-]$ component. As well known, as deformation become larger, this Nilsson
orbit has large contamination of $p_{3/2}$. Final explanation is the quenching of the $N=28$ shell
gap. It has been discussed that the quenching of the $N=28$ shell gap also strongly affects the
neutron-rich Ne and Mg isotopes in the island of inversion where $N=20$ shell gap is broken. A
well known famous example is neutron-halo nucleus $^{31}{\rm Ne}$ with $N=21$, in which the ground
state has $\nu p_{3/2}$ configuration instead of $\nu f_{7/2}$. Even in the case of 
$^{26}{\rm Ne}$ which is out of the island of inversion, the quenching of $N=28$ shell gap will
affect the excitation spectra. Indeed, it is reminded that the $3/2^-$ state is lower than the
$7/2^-$ state in $^{25}{\rm Ne}$. 

\subsection{Isoscalar component of PDR}
The dominance of the core excitation discussed above implies that the PDR has large isoscalar (IS)
dipole strength as well as the IV strength. This is explained as follows. The first line of the
Eq. (\ref{eq:isd1}) is the standard definition of the IS dipole transition operator in terms of
the single-particle coordinate $\bm r_i$ and the center-of-mass coordinate 
$\bm r_{cm}=\sum_{i=1}^A\bm r_i/A$. 

\begin{figure}[h!]
 \includegraphics[width=0.5\hsize]{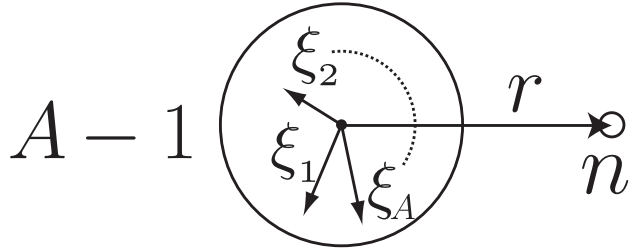}
 \caption{Schematic figure for the internal coordinates $\bm \xi_i$ and the relative coordinate
 $\bm r$.}\label{fig:illust1} 
\end{figure}

\begin{widetext}
\begin{align}
 \mathcal{M}_\mu({\rm IS1} )&=\sum_{i=1}^{A}(\bm r_i - \bm r_{cm})^2
 \mathcal Y_{1\mu}(\bm r_i- \bm r_{cm})\nonumber\\
 &= \sum_{i=1}^{A-1}\xi_i^2\mathcal Y_{1\mu}(\bm \xi_i)
 +\frac{(A-1)(A-2)}{A^2} r^2 \mathcal{Y}_{1\mu}(\bm r)
 - \frac{5}{3A}\sum_{i=1}^{A-1}\xi_i^2\mathcal Y_{1\mu}(\bm r)
 + \frac{4\sqrt{2\pi}}{3A}\left[
 \sum_{i=1}^{A-1}\mathcal Y_{2}(\bm \xi_i)\otimes\mathcal Y_{1}(\bm r)\right]_{1\mu}\nonumber\\
 &\simeq  \frac{4\sqrt{2\pi}}{3A}\left[
 \sum_{i=1}^{A-1}\mathcal Y_{2}(\bm \xi_i)\otimes\mathcal Y_{1}(\bm r)\right]_{1\mu}.
 \label{eq:isd1}
\end{align}
\end{widetext}
Then, we divide the system into the core nucleus with mass $A-1$ and the valence neutron, and
introduce the internal coordinate of the core $\bm \xi_i$ and the relative coordinate between the
core and the valence neutron $\bm r$ (see Fig. \ref{fig:illust1}), 

\begin{align}
 &\bm \xi_i = \bm r_i - \frac{1}{A-1}\sum_{i=1}^{A-1}\bm r_i, \quad i=1,2,...,A-1, \\
 &\bm r = r_A - \frac{1}{A-1}\sum_{i=1}^{A-1}\bm r_i.
\end{align}
Using these coordinates, the operator is equivalently rewritten as the second line of
Eq. (\ref{eq:isd1}) \cite{Chiba2016}. Now we examine each term of the second line. The first term
is the IS dipole excitation of the core nucleus and should have only negligible contribution to
PDR, because it involves the change of the core density, and hence, it cannot contribute to the
low-energy excitation modes. The second term is the dipole excitation of the relative motion
between the core and the valence neutron. The third term is also the dipole excitation of the
relative motion, but it is coupled to the monopole operator of the core. We can also expect that
these two terms cancel out to each other and are negligible. To elucidate it, let us simplify the
wave functions of the  ground state and PDR as, 
\begin{align}
 \ket{{\rm GS}} &= \ket{\Phi_C}\ket{\phi_n},\\
 \ket{{\rm PDR}} &= \sqrt{1-c^2}\ket{\Phi_C}\ket{\phi_n^*} + c\ket{\Phi_C^*}\ket{\phi_n^{**}},
\end{align}
where $\ket{\Phi_C}$ and $\ket{\Phi_C^*}$ are the ground and excited states wave functions of the
core.  $\ket{\phi_n}$ is the valence neutron in the ground state, while $\ket{\phi_n^*}$  and
$\ket{\phi_n^{**}}$ are those in the PDR coupled to the ground and excited states of the core. 
The antisymmetrization between the core and the valence neutron is neglected
for simplicity. Using these wave functions, we estimate the IS dipole transition matrix between
the ground state and PDR. The second term yields 
\begin{align}
 \frac{(A-1)(A-2)}{A^2}\sqrt{1-c^2}\braket{\phi_n^*|r^2\mathcal Y_{1\mu}(\bm r)|\phi_n},
\end{align}
and the third term is
\begin{align}
 -\frac{5}{3}\frac{A-1}{A}\sqrt{1-c^2}\braket{r^2_C}
 \braket{\phi_n^*|\mathcal Y_{1\mu}(\bm r)|\phi_n}. 
\end{align}
Here, we assumed that $\ket{\Phi_C}$ and $\ket{\Phi_C^*}$ have different angular
momenta. $\braket{r^2_C}$ denotes the mean-square radius of the core ground state,  
$ \braket{r^2_C} = \braket{\Phi_C|\sum_{i=1}^{A-1}\xi_i^2|\Phi_C}/(A-1)$. If the radius of the
core $\ket{\Phi_C}$ and that of the valence neutron $\ket{\phi_n}$ are almost the same size, 
we may be able to  expect that the matrix elements 
$\braket{\phi_n^*|r^2\mathcal Y_{1\mu}(\bm r)|\phi_n}$  and 
$\braket{r^2_C} \braket{\phi_n^*|\mathcal Y_{1\mu}(\bm r)|\phi_n}$ are the same order of
magnitude. Hence, we expect that the second and third terms largely cancel out to each other for
such situation. However, it must be noted that this expectation is invalid for the halo nuclei in
which the valence neutron has huge radius. For halo nuclei, the contribution from the second term
will predominate over other terms, and the halo nuclei should have strong IS dipole mode at small
excitation energy. Indeed, this was already pointed out for $^{6}{\rm He}$ \cite{Mikami2014}
theoretically, and recently observed in $^{11}{\rm Li}$ \cite{Kanungo2015}. 

\begin{figure}[h!]
 \includegraphics[width=\hsize]{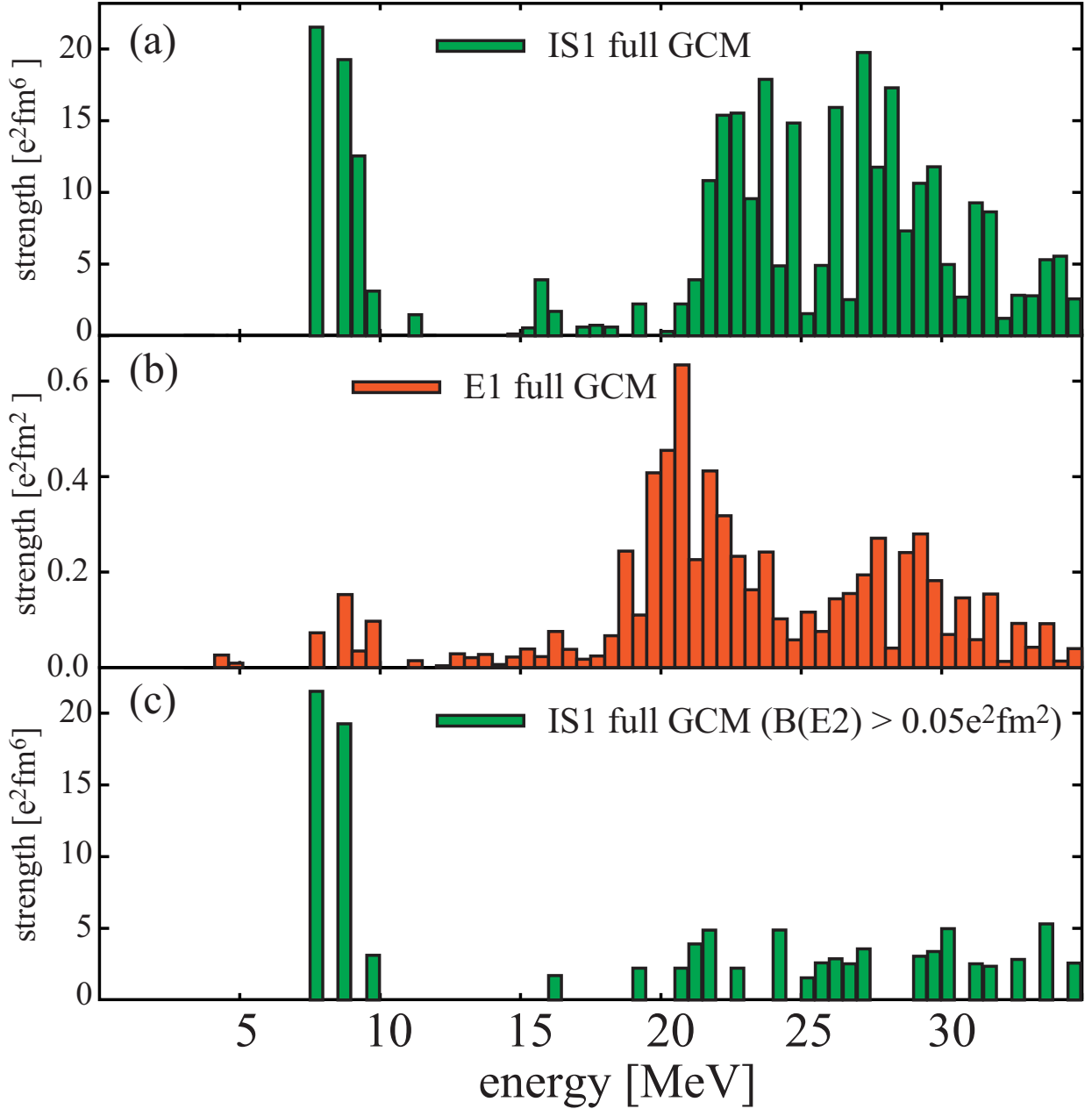}
 \caption{(a) calculated IS dipole strength distribution. (b) calculated electric dipole strength 
 distribution which is the same with Fig. \ref{fig:strength} (a).  (c) calculated IS dipole
 strengths of the states having sizable magnitude of $E1$ strengths 
 $(B(E1)>0.05\ \rm e^2fm^2)$}\label{fig:isd}  
\end{figure}
In the case of the non-halo nucleus $^{26}{\rm Ne}$, we expect that only the forth term has the
sizable contribution to the low-lying dipole mode as written in the last line of
Eq. (\ref{eq:isd1}). It is the dipole excitation of the valence neutron coupled to the
quadrupole operator of the core. Assuming that the ground state of the core has no quadrupole
moment (this is true for the $1/2^+$ ground state of $^{25}{\rm Ne}$), the contribution from
the forth term is estimated as,
\begin{align}
  \frac{4\sqrt{2\pi}}{3A}c \braket{Q_C}\braket{\phi_n^{**}|r\mathcal Y_{1\mu}(\bm r)|\phi_n},
\end{align}
where the quadrupole matrix element of the core is defined as 
$\braket{Q_C}=\braket{\Phi_C^*|\sum_{i=1}^A \mathcal Y_{2\mu}(\bm \xi_i)|\Phi_C}$. Remember that
the first and second excited states of $^{25}{\rm Ne}$ have large $B(E2)$ value
(Tab. \ref{tab:e2}). Hence the matrix element $\braket{Q_C}$ should be large and the fourth term
should yield large IS dipole transition matrix. In other words, the IS dipole transition is
sensitive to the quadrupole excitation of the core, and the PDR of $^{26}{\rm Ne}$ should have
large IS dipole transition matrix if the core excited component is important as discussed in
Sec. \ref{sec:4b}.

This expectation is verified by the numerical calculation of IS dipole strength shown in
Fig. \ref{fig:isd} (a). We clearly see that the PDR has pronounced IS dipole strength as
expected. To make the argument more visible, Fig. \ref{fig:isd} (c) shows the IS dipole
strength of the excited state that have the sizable $E1$ strengths $(B(E1) > 0.05 \rm e^2fm^2)$. 
It is obvious that the PDR has large $E1$ and IS dipole strengths simultaneously, while the other
excited states are not. Thus, the IS dipole strength is correlated well with the core excitation
of PDR and will provide a good insight to the structure of PDR, if experimentally measured.

Knowing above-mentioned results, one may also able to conjecture as follows. Imagine that the PDR
is dominated by the neutron single-particle excitation and proton excitation plays only a minor
role. In such cases, the PDR is not an eigenmode of the isospin, but a mixture of the IV and IS
components, 
\begin{align}
 \ket{\rm PDR} \propto \mathcal M(E1)\ket{\rm GS} + \mathcal M(IS1)\ket{GS}. 
\end{align}
Indeed, this kind of the contamination of the isoscalar component has already been discussed by
many authors \cite{Paar2009,Vretenar2012,Roca-Maza2012,Nakada2013}.
Then, suppose that the core nucleus has strong low-lying quadrupole collectivity. From above
discussions, we can expect that $\mathcal M(IS1)\ket{GS}$ is strongly amplified, and as a result, 
the PDR is predominated by the core excited component. In short, I conjecture that the PDR will be
dominated by the core excited component, if the core nucleus has low-lying strong quadrupole
collectivity. A good candidate of this conjecture is neutron-rich Ne isotopes which have very
strong quadrupole collectivity owing to the breakdown of $N=20$ magic number in the island of
inversion. This conjecture will be tested by the undergoing numerical calculations. 

\section{Summary}
In summary, we have investigated the pygmy dipole resonance of $^{26}{\rm Ne}$ by using the 
shifted-basis AMD. The ordinary AMD framework, $\beta$ GCM, reasonably described the low-lying
spectra of $^{24,25,26}{\rm Ne}$, but failed to describe the $E1$ response of $^{26}{\rm Ne}$. 
The shifted-basis AMD introduces various $1p1h$ configurations by the shift of the nucleon wave
packets and is able to describe $E1$ response.  The global feature of the calculated $E1$ response
function was consistent with the QRPA calculations which employ the same Gogny D1S interaction. It 
is also showed that the splitting of the GDR originates in the ground state deformation.  

The shifted-basis AMD showed that the PDR appears approximately at 8.5 MeV and exhausts XX\% of
the TRK sum which are consistent with the observation. The structure of the PDR was examined by
the analysis of the spectroscopic factors. It was found that the PDR is dominated by the neutron
excitation coupled to the quadrupole excited core nucleus  $^{25}{\rm Ne}$, which explains the
observed decay of PDR to the excited states of  $^{25}{\rm Ne}$. We suggested that the quadrupole 
core excitation induces the large contamination of the isoscalar component in PDR. It was shown by
the analytic calculation by rewriting the isoscalar dipole operator in terms of the internal
coordinates and the relative coordinate between the core and the valence neutron. This estimation
was confirmed by the numerically calculation using shifted-basis AMD. From this result, we
conjecture that the PDR will be
dominated by the core excited component, if the core nucleus has low-lying strong quadrupole
collectivity. By the undergoing numerical calculations, this conjecture will be tested in
neutron-rich Ne isotopes in which the low-lying strong quadrupole collectivity is well known.

\begin{acknowledgments}
The author acknowledge the fruitful discussions with Dr. Kanada-En'yo. He also acknowledges the
support by  the Grants-in-Aid for Scientific Research on Innovative Areas from MEXT (Grant
No. 2404:24105008) and JSPS KAKENHI Grant No. 16K05339. Part of the numerical calculations were
performed by using the super computer in Yukawa Institute for Theoretical Physics (YITP) in Kyoto
University. 
\end{acknowledgments}

\appendix
\section{Overlap amplitude and single-particle spectroscopic factor of Slater determinants} 
\label{sec:appa}
In this appendix, we derive the equations to calculate the overlap amplitude and spectroscopic
factor. The following equations are applicable for nuclear models based on Slater determinant wave 
functions such as Hartree-Fock as well as AMD.

We first consider the Slater determinant wave functions of $A$ and $A+1$ body systems given as,
\begin{align}
 &\Phi(\bm r_1,...,\bm r_A) = \frac{1}{\sqrt{A!}}\det\Set{\phi_1\cdots\phi_{A}},\\
 &\Psi(\bm r_1,...,\bm r_{A+1}) = \frac{1}{\sqrt{(A+1)!}}\det\Set{\psi_1\cdots\psi_{A+1}}.
\end{align}
Using the $A\times (A+1)$ overlap matrix $B_{ij}=\braket{\phi_i|\psi_j}$ and its submatrix
$B^{(p)}$ formed by removing the $p$th column from $B$, the overlap amplitude is calculated as 
\begin{align}
 &\varphi(\bm r)\equiv \sqrt{A+1}\Braket{\Phi|\Psi}\nonumber\\
 &=\sum_{p_1,...,p_{A+1}}sgn
 \left(
 \begin{array}{ccc}
  1,...,A+1\\
 p_1,...,p_{A+1}\end{array}
\right)B_{1p_1}...B_{Ap_A}\psi_{p_{A+1}}(\bm r)\nonumber\\
 &=\sum_{p=1}^{A+1}(-)^{p}\det B^{(p)}\psi_p(\bm r), \label{eq:appa1}
\end{align}
where a trivial factor $(-)^{A+1}$ is omitted for simplicity.
Using this result, we consider the overlap amplitude of angular momentum projected Slater
determinants $P^{J}_{MK}\Phi$ and $P^{J}_{MK}\Psi$. Their overlap amplitude is calculated as
follows. 

\begin{widetext}
\begin{align}
 \varphi(\bm r) &\equiv \sqrt{A+1}\braket{P^{J_2}_{M_2K_2}\Phi|P^{J_1}_{M_1K_1}\Psi}
 \nonumber\\
 &=\frac{(2J_1+1)(2J_2+1)}{(8\pi^2)^2}\int d\Omega_1 d\Omega_2 
 D^{J_2}_{M_2K_2}(\Omega_2) D^{J_1*}_{M_1K_1}(\Omega_1)
 \sqrt{A+1}\braket{R_A(\Omega_2)\Phi|R_{A+1}(\Omega_1)\Psi}\nonumber\\
 &=\frac{(2J_1+1)(2J_2+1)}{(8\pi^2)^2}\sum_K\int d\Omega_1 d\Omega_2' 
 D^{J_2}_{M_2K}(\Omega_1) D^{J_2}_{KK_2}(\Omega_2') D^{J_1*}_{M_1K_1}(\Omega_1)
 \sqrt{A+1}\braket{\Phi|R^\dagger_A(\Omega_2')R^\dagger_A(\Omega_1)
 R_{A+1}(\Omega_1)|\Psi},\label{eq:appa2}
\end{align}
\end{widetext}
where  $\Omega_2'$ satisfies the relation $R(\Omega_2)=R(\Omega_1)R(\Omega_2')$, and hence,
$D^{J_2}_{M_2K_2}(\Omega_2)=\sum_K D^{J_2}_{M_2K}(\Omega_1)D^{J_2}_{KK_2}(\Omega_2')$.
Note that $R_A(\Omega)$ rotates $\bm r_1,...,\bm r_A$, while $R_{A+1}(\Omega)$ rotates $\bm
r_1,...,\bm r_{A+1}$. Then, using Eq. (\ref{eq:appa1}), 
the braket in the integral is calculated. 
\begin{align}
 &\sqrt{A+1}\braket{\Phi|R^\dagger_A(\Omega_2')R^\dagger_A(\Omega_1)
 R_{A+1}(\Omega_1)|\Psi}\nonumber\\
 &=\sum_{p=1}^{A+1}(-)^{p}\det B^{(p)}(-\Omega_2')
 \left\{R(\Omega_1)\psi_{p}(\bm r_{A+1})\right\}.\label{eq:appa3}
\end{align}
Here, $B^{(p)}(-\Omega_2')$ is a $A\times A$ submatrix of
$B_{ij}(-\Omega_2')=\braket{\phi_i|R^\dagger(\Omega_2)|\psi_j}$. 

Now using the multipole expansion,
\begin{align}
 \psi_p(\bm r) = \sum_{jlk}\psi^{(p)}_{jlk}(r)\left[Y_{l}(\hat r)\otimes\chi\right]_{jk},
\end{align}
the rotation of $\psi_p (\bm r)$ is written as
\begin{align}
 R(\Omega_1)\psi_p(\bm r) = \sum_{jlmk}\psi^{(p)}_{jlm}(r) D^{j}_{mk}(\Omega_1)
 \left[Y_{l}(\hat r)\otimes\chi\right]_{jm}.\label{eq:appa4}
\end{align}
Substituting Eqs. (\ref{eq:appa3}) and (\ref{eq:appa4}) into Eq. (\ref{eq:appa2}), the integral
over $\Omega_1$ is analytically performed. Simplifying the equation, we obtain the overlap
amplitude for the angular momentum projected Slater determinants.
\begin{widetext}
\begin{align}
 \varphi(\bm r) &= \sqrt{A+1}\braket{P^{J_2}_{M_2K_2}\Phi|P^{J_1}_{M_1K_1}\Psi}
 =\sum_{jl}C^{J_1M_1}_{J_2M_2,jM_1-M_2} \varphi_{jl}(r) [Y_l(\hat r)\otimes \chi]_{jM_1-M_2},
 \label{eq:appa5}\\
 \varphi_{jl}(r)&=\sum_k C^{J_1K_1}_{J_2K_1-k,jk}\sum_{p=1}^{A+1}(-)^{p}\psi^{(p)}_{jlk}(r)
 \frac{2J_2+1}{8\pi^2}\int d\Omega\ D^{J_2*}_{K_2K_1-k}(\Omega)\det B^{(p)}(\Omega),
 \label{eq:appa6}
\end{align}
\end{widetext}
where $C^{JM}_{j_1m_1,j_2m_2}$ denotes Clebsch-Gordan coefficient. It is obvious that the 
the overlap amplitude of the GCM wave functions given in Eq. (\ref{eq:gcmwf0}), (\ref{eq:gcmwf1})
and (\ref{eq:gcmwf2}) are obtained by a linear transformation of Eq. (\ref{eq:appa6}). 

By a similar manner calculation, the equation for two-body overlap amplitude
$\varphi(\bm r_1,\bm r_2)$ for $A$ and $A+2$ body systems is also obtained as follows.
\begin{widetext}
 \begin{align}
 &\varphi(\bm r_{1},\bm r_{2}) \equiv \sqrt{(A+1)(A+2)}
  \braket{P^{J_2}_{M_2K_2}\Phi|P^{J_1}_{M_1K_1}\Psi}\nonumber\\
  &\qquad\ \ \quad =\sum_{j}C^{J_1M_1}_{J_2M_2,jM_1-M_2}\sum_{j_1l_1j_2l_2}
  \varphi_{j;j_1l_1j_2l_2}(r_1,r_2)
  \left[\left[Y_{l_1}(\hat r_1)\otimes\chi_1\right]_{j_1}\otimes
  \left[Y_{l_2}(\hat r_2)\otimes\chi_2\right]_{j_2}\right]_{jM_1-M_2},\\
  &\varphi_{j;j_1l_1j_2l_2}(r_1,r_2)=\sum_kC^{J_1K_1}_{J_2K_1-k,jk}\sum_{p<q}^{A+2}(-)^{p-q}
  \varphi^{(p,q)}_{jk;j_1l_1j_2l_2}(r_1,r_2)
  \frac{2J_2+1}{8\pi^2}\int d\Omega\ D^{J_2*}_{K_2,K_1-k}(\Omega)\det B^{(p,q)}(\Omega),\\
  &\varphi^{(p,q)}_{jk;j_1l_1j_2l_2}(r_1,r_2)= \sum_{k_1}C^{jk}_{j_1k_1,j_2k-k_1}
  \left\{\psi^{(p)}_{j_1l_1k_1}(r_1)\psi^{(q)}_{j_2l_2k-k_1}(r_2)
  -\psi^{(q)}_{j_1l_1k_1}(r_1)\psi^{(p)}_{j_2l_2k-k_1}(r_2)\right\},
 \end{align}
\end{widetext}
where $B^{(p,q)}(\Omega)$ is a $A\times A$ submatrix which is formed by removing $p$ and $q$
columns from $A\times (A+2)$ matrix $B_{ij}(\Omega)=\braket{\phi_i|R(\Omega)|\psi_j}$. A similar
formula for two-body overlap function was also derived in Ref. \cite{Kobayashi2016}.

\section{Shift of Gaussian centroids\\ and electric dipole operator}\label{sec:appb}
Here, we briefly explain why the shift of the Gaussian wave packets can efficiently describe
various $1p1h$ configurations which coherently contribute the electric dipole modes. The meaning
of the shifting the Gaussian centroid becomes clear when one rewrite the Gaussian wave packet
given in Eq. (\ref{eq:singlewf}) as the coherent state, 
\begin{align}
 \varphi_i({\bf r};\bm Z) &= \prod_{\sigma=x,y,z}
 \left(\frac{2\nu_\sigma}{\pi}\right)^{\frac{1}{4}}
 e^{-\nu_\sigma\left(r_\sigma - 
  \frac{Z_{i\sigma}}{\sqrt{\nu_\sigma}}\right)^2+\frac{1}{2}Z^2_\sigma}\chi_i\xi_i,\nonumber\\
 &=\braket{\bm r|e^{-{Z^2}/{2}}e^{\bm Z\cdot \hat{\bm a}^\dagger}|0}\chi_i\xi_i
 =\braket{\bm r|\bm Z}\chi_i\xi_i,
\end{align}
where $\hat{\bm a}^\dagger=(\hat{a}_x^\dagger,\hat{a}_y^\dagger,\hat{a}_z^\dagger)$ is the
creation operator of the harmonic oscillator with $\hbar\omega_\sigma={2\hbar^2\nu_\sigma}/{m}$
and $\bm Z=(Z_x,Z_y,Z_z)$. The shift of the centroid, $\bm Z \rightarrow \bm Z + \Delta \bm Z$,
is written as
\begin{align}
 \varphi_i({\bf r};\bm Z+\Delta\bm Z) 
 &=\braket{\bm r|e^{-{(Z+\Delta Z)^2}/{2}}e^{(\bm Z+\Delta\bm Z)\cdot
 \hat{\bm a}^\dagger}|0}\chi_i\xi_i\nonumber\\
 &\propto\braket{\bm r|e^{\Delta\bm Z\cdot \hat{\bm a}^\dagger}|\bm Z}\chi_i\xi_i
 =e^{\Delta\bm Z\cdot \hat{\bm a}^\dagger}\phi_i(\bm r;\bm Z).
\end{align}
Thus, by the shift of the centroid, the wave packets are coherently excited, and when 
$\Delta\bm Z$  is sufficiently small, it becomes a linear combination of 0 and $1\hbar\omega$
excitations from the original wave packet. Therefore, when one of the wave packets of a Slater
determinant is slightly shifted, it corresponds to the $1\hbar\omega$ excitation from the original
Slater determinant.

The shift is also closely related to the dipole response. Suppose that the dipole resonances are
well approximated by the ground state wave function multiplied by the $E1$ operator, then it is
rewritten as follows,
\begin{align}
 \ket{E1\ \rm resoance} &\simeq \sum_{i=1}^Z \mathcal Y_{1\mu}(\hat r_i)\ket{GS}\nonumber\\
 &\simeq \frac{1}{\delta}\sum_{i=1}^Z(e^{\delta \mathcal Y_{1\mu}(\hat r_i)}-1)\ket{GS}.
\end{align}
Here $\delta$ is assumed to be sufficiently small number. If $\ket{GS}$ is a Slater determinant
of the Gaussian wave packets, $\sum_{i=1}^{Z}e^{\delta \mathcal Y_{1\mu}(\hat r_i)}\ket{GS}$ may be rewritten
as,
\begin{align}
 &\sum_{i=1}^Ze^{\delta \mathcal Y_{1\mu}(\hat r_i)}\mathcal
 A\set{\varphi_1,...,\varphi_A}\nonumber\\
 &=\sum_{i=1}^Z
 \mathcal A\set{\varphi_1,...,e^{\delta \mathcal Y_{1\mu}(\hat r)}\mathcal \varphi_i,...\varphi_A},
\end{align}
and here $e^{\delta \mathcal Y_{1\mu}(\hat r)}\mathcal \varphi$ corresponds to the shift of the
centroid. For example, in the case of $\mu=0$, it corresponds to the shift along $z$ axis as
follows. 
\begin{align}
 e^{\delta \mathcal Y_{1\mu}(\hat r)}\mathcal \varphi(\bm r;\bm Z_i) &=
 e^{\sqrt{{4\pi}/{3}}\delta z}\mathcal \varphi(\bm r;\bm Z_i) \nonumber\\
 &\propto \varphi(\bm r;\bm Z +\epsilon\bm e_{z}), \\
 \epsilon&=\sqrt{\frac{4\pi}{3}}\frac{\delta}{2\sqrt{\nu}}.
\end{align}
Thus, the dipole modes with small amplitude corresponds to the shift of the Gaussian wave packets
and it generates various $1p1h$ configurations.

\bibliography{Ne26}
\end{document}